\documentclass[aps,prd,final,letterpaper, twocolumn]{revtex4}
\usepackage{amsmath, amssymb, amsthm, cancel, dsfont, enumerate, epstopdf, eucal, fancyhdr, feynmf, gensymb, graphicx, lipsum, mathtools, physics, tikz}
\usepackage[caption=false]{subfig}

\begin{document}

\title{Resonant Neutrino Self-Interactions}
\author{Cyril Creque-Sarbinowski}
\email{creque@jhu.edu}
\affiliation{Department of Physics and Astronomy, Johns Hopkins University, 3400 N. Charles St., Baltimore, MD 21218, USA}
\author{Jeffrey Hyde}
\email{jhyde@bowdoin.edu}
\affiliation{Department of Physics and Astronomy, Bowdoin College, 8800 College Station, Brunswick, ME 04011-8488, USA}
\author{Marc Kamionkowski}
\email{kamion@jhu.edu}
\affiliation{Department of Physics and Astronomy, Johns Hopkins University, 3400 N. Charles St., Baltimore, MD 21218, USA}
\date{\today} 

\begin{abstract}
If neutrinos have self-interactions, these will induce scatterings between astrophysical and cosmic neutrinos. Prior work proposed to look for possible resulting resonance features in astrophysical neutrino spectra in order to seek a neutrino self-interaction which can be either diagonal in the neutrino flavor space or couple different neutrino flavors.  The calculation of the astrophysical spectra involves either a Monte Carlo simulation or a computationally intensive numerical integration of an integro-partial-differential equation.  As a result only limited regions of the neutrino self-interaction parameter space have been explored, and only flavor-diagonal self-interactions have been considered.  Here, we present a fully analytic form for the astrophysical neutrino spectra for arbitrary neutrino number and arbitrary self-coupling matrix that accurately obtains the resonance features in the observable neutrino spectra.  The results can be applied to calculations of the diffuse supernova neutrino background and of the spectrum from high-energy astrophysical neutrino sources.  We illustrate with a few examples.
\end{abstract}

\maketitle

\pagestyle{myheadings}
\markboth{Cyril Creque-Sarbinowski}{Resonant Neutrino Self-Interactions}
\thispagestyle{empty}

\section{Introduction}
While the interactions between neutrinos is extraordinarily feeble in the Standard Model, there are a number of reasons to entertain the possibility that new physics may introduce stronger neutrino self-interactions~\cite{1409.4180, 1508.07471, 1903.02036, 1911.06342}.  Astrophysical neutrinos may provide a powerful probe in the search for such self-interactions ($\nu$SI)~\cite{1907.00991}. Strong features, such as dips and enhancements, can be imprinted on astrophysical-neutrino spectra, which when analyzed can yield $\nu$SI parameter values.  In particular, there is the diffuse supernova neutrino background (DSNB) and a collection of high-energy astrophysical neutrino (HEAN) sources. The DSNB is the isotropic time-independent flux of neutrinos and anti-neutrinos around tens of MeVs emitted by distant core-collapse supernovae~\cite{1004.3311}. These diffuse sources come from distances around 10 Mpc~\cite{astro-ph/0503321, 1304.2553} up to a redshift of 5 with a peak around a redshift of 1~\cite{0812.3157}. In comparison, although HEAN sources have no identified production mechanism, they are observed by IceCube to follow a power law~\cite{2001.09520}. 

If there are neutrino self-interactions, then interactions of DSNB and/or HEAN neutrinos with low-energy ($\sim0.001$~eV) cosmic-background  neutrinos can appear in the obseved DSNB and/or HEAN spectra as absorption features or enhancements at lower energies.  The calculation of the observed flux is straightforward but, most generally, involves solving a series of coupled integro-differential equations that evolve the energy and flavor distribution of neutrinos.  These equations describe the injection of neutrinos from sources, the redshifting of neutrinos, the absorption from self-interactions, and the reinjection of lower-energy neutrinos after such interactions.   The solutions to these equations require either a Monte-Carlo simulation or a straightforward, but computationally intensive, numerical integration of the equations.  This thus limits the regions of $\nu$SI parameter space that can be investigated.  For example, previous work assumed a universal self-coupling matrix~\cite{1912.09115, hep-ph/0607281, hep-ph/0404221}, a diagonal~\cite{1408.3799} self-coupling matrix, no flavor dynamics~\cite{1404.2288, 2001.04994}, or particular values for the mediator mass, coupling constant, and neutrino mass~\cite{1401.7019, 1407.2848, 1803.04541}.

In this paper, we present an analytic approach to resonant astrophysical-cosmic neutrino scattering for arbitrary self-coupling matrix and neutrino number. This solution is built on the observation that most of observable effects explored in DSNB/HEAN studies arise from resonant neutrino self-interactions.  To illustrate the utility of the approach, we then use this solution to explore the discovery space for a model of  $\tau$-self-interactions (relevant for the DSNB) and another for sterile-neutrino self-interactions (relevant for the HEAN).

This paper is organized as follows. In Section~\ref{sec:formalism}, we review the general formalism of neutrino mixing and transport. In doing so, we present three formal solutions: first in the case of no interactions; second in the case of only absorption interactions; and third in the case of both absorption and reinjection. We formulate explicit solutions in Section~\ref{sec:analytic} where we specify the nature of neutrino self-interaction. We begin by considering only a single species of neutrino. Then, we generalize this result and present a solution for arbitrary neutrino number and self-coupling matrix. Finally, in Section~\ref{sec:numeric} we use our new solution to identify regions of parameter space that can be accessed. Specifically, we consider DSNB probes of self-interactions with Super-Kamiokande and HEAN probes with IceCube. We discuss and conclude these results in Section~\ref{sec:disc} and Section~\ref{sec:conc}.

\section{General Formalism}\label{sec:formalism}
\subsection{Neutrino Mixing}\label{subsec:mixing}
Neutrinos can be represented in either the mass basis or the flavor basis. In what follows, Greek indices are reserved for flavor states and Latin indices for mass states. In order to switch between the two bases, the neutrino mixing matrix $U$ must be used according to
\begin{align}
\nu_\alpha = \sum_i U_{\alpha i} \nu_i,
\end{align}
where the sum is over all mass states and with $U$ unitary. Unitarity implies that for any flavor state $\alpha$, $\sum_{i} |U_{i\alpha}|^2 = 1$, and so $|U_{\alpha i}|^2$ is interpreted as the probability that flavor $\alpha$ is observed as mass state $i$, or {\it vice verse}. 

\subsection{Neutrino Transport}\label{subsec:transport}
The specific flux $\Phi_i(t, E)$ of astrophysical neutrinos $\nu_i$ (number of astrophysical neutrinos per unit conformal time per unit comoving area per unit energy) at cosmic time $t$ and observed energy $E$ obeys the Boltzmann equation
\begin{align}\label{eq:boltzmann_flux}
\nonumber \frac{\partial \Phi_i}{\partial t} &= H\Phi_i + HE\frac{\partial \Phi_i}{\partial E} + S_i(t, E)\\
& - \Gamma_i(t, E)\Phi_i+ S_{{\rm tert}, i}(t, E),
\end{align} 
where $H(z) = H_0 \widetilde{E}(z)$ is the Hubble parameter at redshift $z$, with $z$ as a proxy for $t$, and for $\Lambda$CDM, $\widetilde{E}(z) = [\Omega_m(1 + z)^3 + (1 - \Omega_m)]^{1/2}$, with $\Omega_m$ the matter-density parameter today. Moreover, $S_i$ is the production rate of astrophysical neutrinos $\nu_i$, $\Gamma_i$ is the absorption rate of astrophysical neutrinos due to neutrino scatterings, and $S_{{\rm tert}, i}$ is the tertiary source term accounting for the possible reinjection of astrophysical neutrinos post-scattering. Note that by this definition of the specific flux, the comoving number density of astrophysical neutrinos is $(1/c)\int dE \ \Phi_i(t, E)$, different from Refs.\cite{1312.3501, 1811.04939} where they considered $\Phi_i$ to be defined to reproduce the physical number density. Since neutrino decoherence time scales are smaller than any other relevant time scale, the transformation $\Phi_\alpha(t, E) = \sum_i |U_{\alpha i}|^2 \Phi_i(t, E)$ can be performed to switch back to the flavor basis at any point. Moreover, if the source terms are given in the flavor basis $S_\alpha$, then the mass basis source term is $S_i = \sum_\alpha |U_{\alpha i}|^2 S_\alpha$. 

In the absence of interactions, $S_{{\rm tert}, i} = \Gamma_i = 0$, the solution of Eq.~\eqref{eq:boltzmann_flux} is obtained by identifying the total time derivative as $d/dt = \partial/\partial t + (dE/dt)\partial/\partial E$, with $dE/dt = -HE$, leading to
\begin{align}\label{eq:noint_flux}
\Phi_i(t, E) &= \int_{-\infty}^t dt' [a(t)/a(t')]S_i\{t', [a(t)/a(t')]E\},
\end{align}
with $a(t)$ the scale factor at time $t$. The factor of $[a(t)/a(t')]$ inside the source term accounts for the redshifting of the energy from $t'$ to $t$, while outside the source term it accounts for the redshifting of the differential energy from $t'$ to $t$. The addition of a nonzero sink term while neglecting any reinjection, $\Gamma_i \neq 0,\ S_{{\rm tert}, i} = 0$, does not complicate things much further as then
\begin{align}\label{eq:absorb_flux}
\nonumber \Phi_i(t,E) &= \int_{-\infty}^{t}dt' [a(t)/a(t')]e^{-\tau_i(t', t, E)} S_i\{t', [a(t)/a(t')]E\},\\
\tau_i(t', t, E) &= \int_{t'}^t dt'' \Gamma_i \{t'', [a(t)/a(t'')]E\},
\end{align}
with $\tau_i(t', t, E)$ the optical depth of a neutrino $\nu_i$ of energy $E$ between times $t'$ and $t$. As a result, astrophysical neutrinos at time $t'$ not only go through the previous redshifting, but now travel through a medium of optical depth $\tau_i$ from the emission time $t'$ to the observed time $t$. 

Formally, if neutrino reinjection is taken into account, $S_{\rm tert} \neq 0$, a solution is easily written down, 
\begin{align}\label{eq:tert_flux}
\nonumber \Phi_i(t,E) &= \int_{-\infty}^{t}dt' [a(t)/a(t')]e^{-\tau_i(t', t, E)} \widetilde{S}_i\{t', [a(t)/a(t')]E\},\\
\nonumber \tau_i(t', t, E) &= \int_{t'}^t dt'' \Gamma_i \{t'', [a(t)/a(t'')]E\},\\
\widetilde{S}_i(t, E) &= S_i(t, E) + S_{{\rm tert}, i}(t, E).
\end{align}
However, since the tertiary source is a function of the specific flux itself, the solution is in general not closed. Therefore, if we are to move forward, a particular model must be specified. 
\section{Analytical Results}\label{sec:analytic}
\subsection{Single Neutrino Species}\label{subsec:single}
The particular neutrino model we consider at first is that of a single species of self-interacting neutrinos $\nu$ of mass $m_\nu$ whose self-interactions are mediated by a scalar particle $\phi$ with mass $m_\phi$ and coupling strength $g$. We ignore the existence of other neutrino species, and as such we suppress any indices present in relevant equations. That is, we initially consider the interacting Lagrangian,
\begin{align}
\mathcal{L}_{\rm int}^{1-\nu} = g\phi \nu \nu. 
\end{align}
If this is the case, then astrophysical neutrinos will scatter with cosmic neutrinos, causing depletion of astrophysical neutrinos at a resonant energy $E_R = [m_\phi^2/(2 m_\nu)]c^2$ at a rate $\Gamma(t, E) = n_\nu(t)\sigma(E) c$. Here and onwards, cosmic neutrinos will refer to cosmic-background neutrinos. We define $n_\nu(t)$ to be the physical number density of our single cosmic neutrino species, $\sigma(E)$ the scattering cross section of the process $\nu\nu\rightarrow\nu\nu$, and $c$ the speed of light. After depletion, neutrinos are then re-injected at energies $E < E_R$. We take the scattering cross-section to have a Breit-Wigner form 
\begin{align}
\frac{\sigma(E)}{(\hbar c)^2} = \frac{g^4}{4\pi}\frac{s}{[s - (m_\phi c)^2]^2 + (m_\phi c^2)^2 \Gamma_\phi^2},
\end{align} 
where $\hbar$ is Planck's constant, $s = 2 E m_\nu c^2$, and $\Gamma_\phi= g^2 m_\phi c^2/(4\pi)$ the decay width. If the width of the resonance is small enough, resonant scattering can be approximated by a Dirac delta function. We now quantify when this occurs. The width of the resonance is where $[s - (m_\phi c^2)^2]^2 < (m_\phi c^2)^2\Gamma_\phi^2$, or stated in terms of energies when $|E  - E_R[1 \pm \Gamma_\phi/(m_\phi c^2)]| < 0$, so that the width is $2 E_R \Gamma_\phi/(m_\phi c^2)$. For a detector with resolution $\Delta E$, a width cannot be resolved and thus is a delta function if $2E_R\Gamma_\phi/(m_\phi c^2) \lesssim \Delta E$. Therefore the coupling must satisfy
\begin{align}
\nonumber g &\lesssim \sqrt{2\pi}(\Delta E/E_R)^{1/2}\\
&\lesssim 0.5 \left[\frac{\Delta E/(1\ {\rm MeV})}{E_R/(25\ {\rm MeV})}\right]^{1/2},
\end{align}
where $\Delta E \approx 1\ {\rm MeV}$ for a detector such as Super-K~\cite{1111.5031}, and $E_R \approx 25\ {\rm MeV}$ for masses $m_\phi c^2 = 1\ {\rm keV}$, $m_\nu c^2 = 2\times 10^{-2}\ {\rm eV}$. We conclude that unless the coupling is of order unity, which is most of the available parameter space~\cite{1404.2288}, a detector will not be able to resolve the resonance and we approximate the cross section as a delta function. A nascent delta function in the Breit-Wigner form is $\delta(x) = \lim_{\epsilon \rightarrow 0} (1/\pi)\epsilon/(\epsilon^2 + x^2)$, so that the resulting cross section is
\begin{align}
\sigma(E) &= \sigma_{R}E\delta\left(E - E_R\right),
\end{align} 
with $\sigma_{R} = (\hbar c)^2 \pi g^2/(m_\phi c^2)^2$. Resonant scattering is isotropic when the mediator is a scalar field, and therefore the differential cross section $d\sigma(E_1, E_3)/dE_3$, where an incoming neutrino with energy $E_1$ scatters to an outgoing neutrino of energy $E_3$, has a flat distribution $d\sigma(E_1, E_3)/dE_3 = \sigma(E_1)/E_1$. With this form we now evaluate the tertiary source for neutrino production in Eq.~\eqref{eq:boltzmann_flux}. 

In our case of cosmic neutrino upscattering, two neutrinos are re-injected after an initial neutrino is taken from the sink term, and cosmic neutrinos have energies much smaller than supernova neutrinos so their relative velocity is the speed of light. Thus the tertiary term takes the following expression, converting our initial differential equation into an integro-differential equation
\begin{align}
\nonumber S_{\rm tert}(t, E)= n_{\nu}(t)c&\int_E^\infty dE_1 \Phi(t, E_1)\\
&\times \left[\frac{d\sigma(E_1, E)}{dE} + \frac{d\sigma(E_1, E_1 - E)}{dE}\right].
\end{align}
With our delta-function approximation, this term is now evaluated as
\begin{align}
S_{\rm tert}(t, E) &= 2\Gamma_R(t)\Phi(t, E_R)\Theta(E_R - E), 
\end{align}  
with $\Gamma_{R}(t) = n_{\nu}(t)\sigma_{ R}c$ and $\Theta(x)$ the Heaviside function with $\Theta(0) = 0$. Moreover, we simplify the optical depth as 
\begin{align}
\tau(t', t, E) = \tau_R(t, E)\Theta[z_R(t, E) - z]\Theta[z' - z_R(t, E)],
\end{align}
with $z'$ a proxy for $t'$, $\tau_R(t, E) = [\Gamma_R(z_R)/H(z_R)][(1 + z)/(1 + z_R)]$, and $z_R = (1 + z)E_R/E - 1$ the absorption redshift of a neutrino with energy $E$. Plugging these expressions into Eq.~\eqref{eq:tert_flux} leads to\begin{align}\label{eq:solved_tert_flux}
\nonumber \Phi(t,E) &= \int_{-\infty}^{t}dt'\ [a(t)/a(t')]e^{-\tau(t', t, E)} \widetilde{S}\{t', [a(t)/a(t')]E\},\\
\widetilde{S}(t, E) &= S(t, E) + 2\Gamma_{R}(t)\Phi(t, E_R)\Theta(E_R - E).
\end{align} 
Thus, for $E\geq E_R$ the spectrum is the same as a no-interaction Boltzmann equation, and so is solved in the same manner. However, when $E<E_R$, neutrinos are reinjected at twice the rate of their depletion at the resonant energy. As such, the expression for neutrino reinjection still requires solving for the specific flux at $E = E_R$ and plugging it back in for evaluation at lower energies, which at first makes Eq.~\eqref{eq:solved_tert_flux} seem not closed. However, Eq.~\eqref{eq:boltzmann_flux} has a delta function via the absorption term $\Gamma(t,E)\Phi = n_{\nu}(t)\sigma(E) c \Phi$ at the resonant energy, and so we must obey the boundary condition at this point. In order to satisfy this condition, we integrate Eq.~\eqref{eq:boltzmann_flux} around the resonant energy from below the resonance $E_R^- \equiv E_R - \epsilon/2$ to above the resonance $E_R^+ \equiv E_R + \epsilon/2$, and take the line width $E_R^+ - E_R^- = \epsilon$ to zero. Explicitly, this results in 
\begin{align}
H(t)\left[\Phi(t, E_R^+) - \Phi(t, E_R^-)\right] &= \Gamma_R(t)\Phi(t, E_R).
\end{align}
Again, above the resonant line the optical depth of free-streaming with no interactions is zero, while below, it is $\tau(t', t, E_R^-)$, so that the resulting expression for $\Phi(t, E_R)$ is
\begin{align}\label{eq:res_flux}
\nonumber \Phi(t, E_R) = \frac{H(t)}{\Gamma_R(t)}\int_{-\infty}^t dt' &[a(t)/a(t')]\left[1 - e^{-\tau_R(t, E)}\right]\\
\times & S\{t', [a(t)/a(t')]E_R\}.
\end{align}
As a result, Eq.~\eqref{eq:solved_tert_flux} has a closed form expression. The purpose of retaining $E_R^-$ in this expression is as a reminder that in order to evaluate $\tau(t', t, E_R^-)$ one must take the left-side limit of the integral in the expression of $\tau$. Since cosmic neutrinos are low-energy, astrophysical-cosmic neutrino scattering does not add or remove energy from the astrophysical neutrino spectra, but only redistributes it. We have checked both analytically and numerically that Eq.~\eqref{eq:solved_tert_flux} obeys this condition.

There are two differences in the expression for $\Phi$ between Eq.~\eqref{eq:solved_tert_flux} and Eq.~\eqref{eq:res_flux}. First, is the presence of the factor $1 - e^{-\tau}$ rather than $e^{-\tau}$. This factor can be understood as follows: after an astrophysical neutrino redshifts through a resonance over a short period of time, the specific flux at the resonant energy only has a fraction $e^{-\tau}$ remaining of the original flux. It follows then that the amount that is injected at lower energies must be the complementary fraction, $1 - e^{-\tau}$. The second difference is the factor of $H(t)/\Gamma_R(t)$, which changes the rate of injection from $\Gamma_R(t)$ in Eq.~\eqref{eq:tert_flux} to $H(t)$. This change in the rate of injection is due the resonance line redshifting in time. If the scattering rate is faster than Hubble, then the flux of neutrinos at the resonant energy is suppressed. Conversely, if the rate is slower, then the scatterings have little effect. 
\subsection{Multiple Neutrino Species}\label{subsec:multi}
In the presence of multiple neutrino species, the previous equations do not hold. Here we present the analogous calculation with additional neutrinos, taking  the mass of each neutrino species to be $m_j$ with corresponding cosmic physical number density $n_j$. The interaction Lagrangian term for the most general mass-basis interaction with a scalar mediator $\phi$ of mass $m_\phi$ is given by
\begin{align}
 \mathcal{L}_{\rm int}^{\rm mass} &= \phi\sum_{i j}g_{i j}\nu_i\nu_j,
\end{align}
with $g_{ij}$ the self-coupling matrix. In this model, astrophysical neutrinos scatter off of one of any of the cosmic neutrino species, causing depletion of astrophysical neutrinos at corresponding resonant energies $E_j = [m_\phi^2/(2m_j)]c^2$ at a rate $\Gamma_i \equiv \sum_j n_j \sigma_{ij} c$. The cross section $\sigma_{ij} \equiv \sum_{kl}\sigma_{ijkl}$ is the sum of scattering cross sections for the processes $\nu_i\nu_j \rightarrow \nu_k\nu_l$. We take $\sigma_{ijkl}$ to have the Briet-Wigner form
\begin{align}
\frac{\sigma_{ijkl}(E)}{(\hbar c)^2} = \frac{|g_{ij}|^2|g_{kl}|^2}{4\pi}\frac{s_j}{[s_j - (m_\phi c)^2]^2 + (m_\phi c^2)^2\Gamma_\phi^2} 
\end{align}
with $s_j = 2E m_jc^2$, and $\Gamma_\phi = \left(\sum_{ij}|g_{ij}|^2\right)m_\phi c^2/(4\pi)$ the decay width. Note that the decay width has changed since there now exists multiple decay branches for $\phi$. After depletion, the neutrinos are re-injected at a rate according to
\begin{align}
\nonumber S_{{\rm tert}, i}(t, E) &= \sum_{jkl}n_k(t)c \int_E^\infty dE_1 \Phi_j(t, E_1)\\
&\times\left[\frac{d\sigma_{jkil}(E_1, E)}{dE} + \delta_{il}\frac{d\sigma_{jkil}(E_1, E_1 - E)}{dE}\right],
\end{align}
with $\delta_{il}$ the Kronecker delta function that accounts for the possibility of upscattering into two astrophysical neutrinos of the same state rather than just one. That is, compared to the single neutrino species case, this expression accounts for production of neutrinos of type $i$ from an astrophysical flux $\Phi_j$ hitting a cosmic neutrino density $n_k$, with $i,j,k$ not necessarily all being the same. Once again, the differential cross section takes a flat distribution $d\sigma_{ijkl}(E_1, E_3)/dE_3 = \sigma_{ijkl}(E_1)/E_1$. Moreover, using our delta function limit, the cross section takes the form $\sigma_{ijkl}(E) = \sigma_R^{ijkl} E\delta(E - E_j)$ with $\sigma_R^{ijkl} = (\hbar c)^2 |g_{i j}|^2|g_{k l}|^2/[4(m_\phi c^2)\Gamma_\phi]$. As a result the tertiary source term is
\begin{align}
\nonumber S_{{\rm tert}, i}(t, E) = \sum_{jkl}\left(1 + \delta_{il}\right)\Gamma_R^{jkil}(t)\Phi_j(t, E_k)\Theta(E_k - E),
\end{align}
with $\Gamma_R^{jkil}(t) = n_k(t)\sigma_R^{jkil}c$.  In addition, the optical depth is
\begin{align}
\tau_i(t', t, E) = \sum_j \tau^{ij}_R(t, E)\Theta[z_j(t, E) - z]\Theta[z' - z_j(t, E)],
\end{align}
with $\tau^{ij}_R(t, E) = [\Gamma^{ij}_R(z_j)/H(z_j)][(1 + z)/(1 + z_j)]$, $\Gamma^{ij} = \sum_{kl}\Gamma_R^{ijkl}$, and $z_j = (1 + z)E_j/E - 1$. Analogous to before, we satisfy the boundary condition around each resonance by the conditions  
\begin{align}
H(t)[\Phi_i(t, E^+_j) - \Phi_i(E_j^-)] = \Gamma^{ij}_R(t)\Phi_i(t, E_j).
\end{align}
Now, it is true that only above the highest resonant line the optical depth is zero. Thus, the general solution is
\begin{align}\label{eq:gen_flux}
\nonumber \Phi_i(t, E) &= \int_{-\infty}^t dt'[a(t)/a(t')]e^{-\tau_i(t', t, E)}\widetilde{S}_i\{t', [a(t)/a(t')E]\},\\
\nonumber \widetilde{S}_i(t, E) &= S_{i}(t, E) + \sum_{jkl}\left(1 + \delta_{il}\right)\Gamma_R^{jkil}(t)\Phi_j(t, E_k)\Theta(E_k - E),\\
\nonumber \Phi_i(t, E_j) &= \frac{H(t)}{\Gamma^{ij}_R(t)}\int_{-\infty}^t dt' \frac{a(t)}{a(t')}e^{-\tau_i(t', t, E)}\left[1 - e^{-\tau^{ij}_R(t, E)}\right]\\
\hphantom{\Phi_i(t, E_j)}&\hphantom{= \frac{H(t)}{\Gamma^{ij}_R(t)}\int_{-\infty}^t dt'\ }\times\widetilde{S}_i\{t', [a(t)/a(t')]E_j\}.
\end{align}
Then when we want to convert back to the flavor basis we use the neutrino mixing matrix once again. Eq.~\eqref{eq:gen_flux} is our main result that describes the propagation of multiple astrophysical neutrinos species that self-interact arbitrarily with cosmic neutrinos. If a flavor self-coupling matrix is given instead, the identification $g_{ij} = \sum_{\alpha\beta}U_{i\alpha}U_{j\beta}g_{\alpha\beta}$ leads to an easy substitution. We present the analogous equations with this substitution in Appendix~\ref{ap:flavor_int}. 

Note however that in order to solve Eq.~\eqref{eq:gen_flux} in a closed manner, the highest resonant boundary condition must be solved for first, as it is a function of only the source $S_i$ and scattering rate $\Gamma_i$. This is to be contrasted with the boundary conditions for lower resonant energies, which depend not only on these quantities, but also the flux at higher resonances. This dependence arises because a neutrino can be absorbed at a resonance $E_j$, downscattered to an energy $E > E_k$, with $E_k < E_j$ some other resonance, and then be redshifted down to $E_k$. In this way, astrophysical neutrinos may cascade down the resonance pipeline until they reach energies below the lowest resonant energy.

\section{Numerical Results}\label{sec:numeric}
Given these analytic results, a wealth of neutrino self-interactions can be explored and constrained. However, due to the large dimensionality of the general problem, we narrow our scope to two specific models. Moreover, many factors aside from neutrino-self interactions can affect the resulting spectrum, such as detector backgrounds and energy thresholds. Such a detailed analysis, however, is outside the scope of this paper. That is, we consider only a single source of neutrinos with shot-noise error.  Specifically, first we consider the standard 3-neutrino model, adjoined with a $\tau$ self-interaction coupling constant $g_{\tau\tau}$. Interactions of this form have been proposed to resolve the Hubble tension \cite{1905.02727}, although our analysis does not rely on this explanation.

Second, we add a sterile neutrino to the 3-neutrino model, along with a sterile $s$ self-interaction coupling constant $g_{ss}$. This case is motivated by the LSND, MiniBooNE and reactor anomalies which suggest mixing with eV-scale sterile neutrinos~\cite{hep-ex/0104049, 1101.2755,1805.12028}. While such mixing would be in tension with Planck, self-interactions of the sterile neutrino by a mediator of mass $ m_\phi \lesssim {\rm MeV}$ would bring results back into harmony~\cite{1310.6337, 1310.5926}. In both models we consider  all other neutrino self-coupling constants are taken to be zero.

In order to apply Eq.~\eqref{eq:gen_flux} to the four-neutrino case, we need to choose definite values for the mixing matrix elements. We use a standard parameterization~\cite{1801.04855}
\begin{align}\label{eq:mixing_matrix}
U  = R^1(3, 4)R^0(2, 4)R^1(1, 4)R^0(2, 3)R^1(1, 3)R^0(1, 2),
\end{align} 
with $R^c(a, b)$ a $4\times 4$ rotation matrix with matrix elements $R^c(a, b)_{ij}$ and a mixing angle $\theta_{ab}$. The matrix elements are those of the $4\times 4$ identity except for the following submatrix:
\begin{align}
\left(\begin{array}{cc} R^c(a, b)_{aa} & R^c(a, b)_{ab}\\ R^c(a, b)_{ba} & R^c(a, b)_{bb}\end{array}\right) &= \left(\begin{array}{cc} c_{ab} & s_{ab}e^{-i c \delta_{ab}^{\rm CP}}\\ -s_{ab}e^{i c \delta_{ab}^{\rm CP}} & c_{ab}\end{array}\right),
\end{align}
where $s_{ab} = \sin(\theta_{ab})$ and $c_{ab} = \cos(\theta_{ab})$, and $\delta_{ab}^{\rm CP}$ is a complex CP violating phase. In general there are also Majorana phases associated with the mixing matrix, but since since we considering lepton-conserving processes, we neglect them~\cite{1710.00715}.

In the limiting case of no mixing between active and sterile neutrino states, $\theta_{14} = \theta_{24} = \theta_{34} = 0$, each of $R^1(3, 4)$, $R^0(2, 4)$, $R^1(1, 4)$ is the identity and we obtain the standard 3-neutrino mixing matrix~\cite{hep-ph/0310238} plus a decoupled sterile state. Note that in this model we consider self-interactions only among sterile neutrinos, so in this no-mixing limit our astrophysical spectra will return to the standard expectation, regardless of the value of $g_{ss}$. Motivated by the short-baseline anomalies, we take $\theta_{14} = \theta_{34} = 0$ and $\sin^2(\theta_{24}) = 0.1$, so $\theta_{24} = 0.161$. 

In addition to the mixing matrix, the neutrino mass spectrum $\vec{m}$ is also constrained. We first review the constraints on the lightest three neutrinos. Oscillation experiments give the value of two mass-squared differences~\cite{1811.05487}. As a result, it is unclear whether the neutrino mass spectrum follows a normal hierarchy (NH) $m_1 < m_2 < m_3$ or inverted hierarchy (IH) $m_3 < m_1 < m_2$. However, a lower bound on the neutrino masses is obtained by setting $m_1 = 0$ in the NH and $m_3 = 0$ in the IH. In addition, an upper bound is obtained from Planck~\cite{1807.06209}, as it constrains the sum of neutrino masses to be such that $\sum_j m_j c^2< 0.12\ {\rm eV}$. As a result, the following table of neutrino mass constraints can be made
\begin{align*}
\begin{tabular}{|c|c|c|c|}
\hline
 & $m_1c^2$ [eV] & $m_2c^2$ [eV] & $m_3c^2$ [eV]\\
\hline
NH & $[0, 0.030]$ & $[0.0087, 0.031]$ & $[0.050,0.059]$\\
IH & $[0.050, 0.052]$ & $[0.051, 0.053]$  & $[0, 0.015]$\\
\hline
\end{tabular}\label{table:m_nu}
\end{align*}
This table implies that no matter the hierarchy, we know there exists a neutrino with mass $m c^2\in [0.050, 0.059]\ {\rm eV}$. Thus, there is at least one cosmic neutrino that is cold today. As an exemplar case of multiple resonances, we choose our three-neutrino mass spectrum to be the heaviest normal hierarchy allowed $\vec{m}_{\rm HNH} c^2 = [0.030, 0.031, 0.059]\ {\rm eV}$. When considering sterile self-interactions, we add to the heaviest normal hierarchy an eV-mass neutrino, leading to a sterile normal hierarchy $\vec{m}_{\rm SNH}c^2 = [0.03, 0.031, 0.059, 1.0]\ {\rm eV}$.
  
The mass of the mediator is chosen to correspond to the energy ranges dictated by the sources we choose. That is, for an experiment that measures spectra between neutrino energy ranges $[E_{\rm min}, E_{\rm max}]$, the range of mediators that can be be probed is $E_{\rm min}\leq [m_\phi^2/(2m_j)]c^2\leq E_{\rm max}$ for each neutrino mass $m_j \in \vec{m}$. In order to simplify our analysis we only compare the null hypothesis with our model at the resonant energies, and not the entire spectrum. Finally, we choose a fixed bin size for each constraint. 

We denote the event count under the null hypothesis $g_{ij} = 0$ by $N_{\rm null}$. Thus, assuming only Poisson shot noise error, we find that the number of events $N_{\rm events}$ in each bin can be measured away from the null hypothesis with a signal to noise
\begin{align}
\left(\frac{S}{N}\right)^2 = \frac{\left(N_{\rm events} - N_{\rm null}\right)^2}{N_{\rm events} + N_{\rm null}}.
\end{align} 
Therefore, the number of events needed to distinguish from the null hypothesis $N_{\rm events} = N_{\rm null}$ is
\begin{align}
N_{\pm} = N_{\rm null} + (S/N)^2/2 + (S/N)_{\pm}\sqrt{2N_{\rm null} + (S/N)^2/4},
\end{align}
with $(S/N)_{\pm} = \pm |(S/N)|$. $N_{\pm}$ is also known as the $(S/N)$-$\sigma$ uncertainty in the measurement of $N_{\rm null}$, with $N_+$ the upper and $N_-$ the lower uncertainty. Again, for our analysis, we only use $N_-$ when looking for depletions.   
\subsection{DSNB}\label{subsec:DSNB}
\begin{figure}[]
\includegraphics[width = \linewidth]{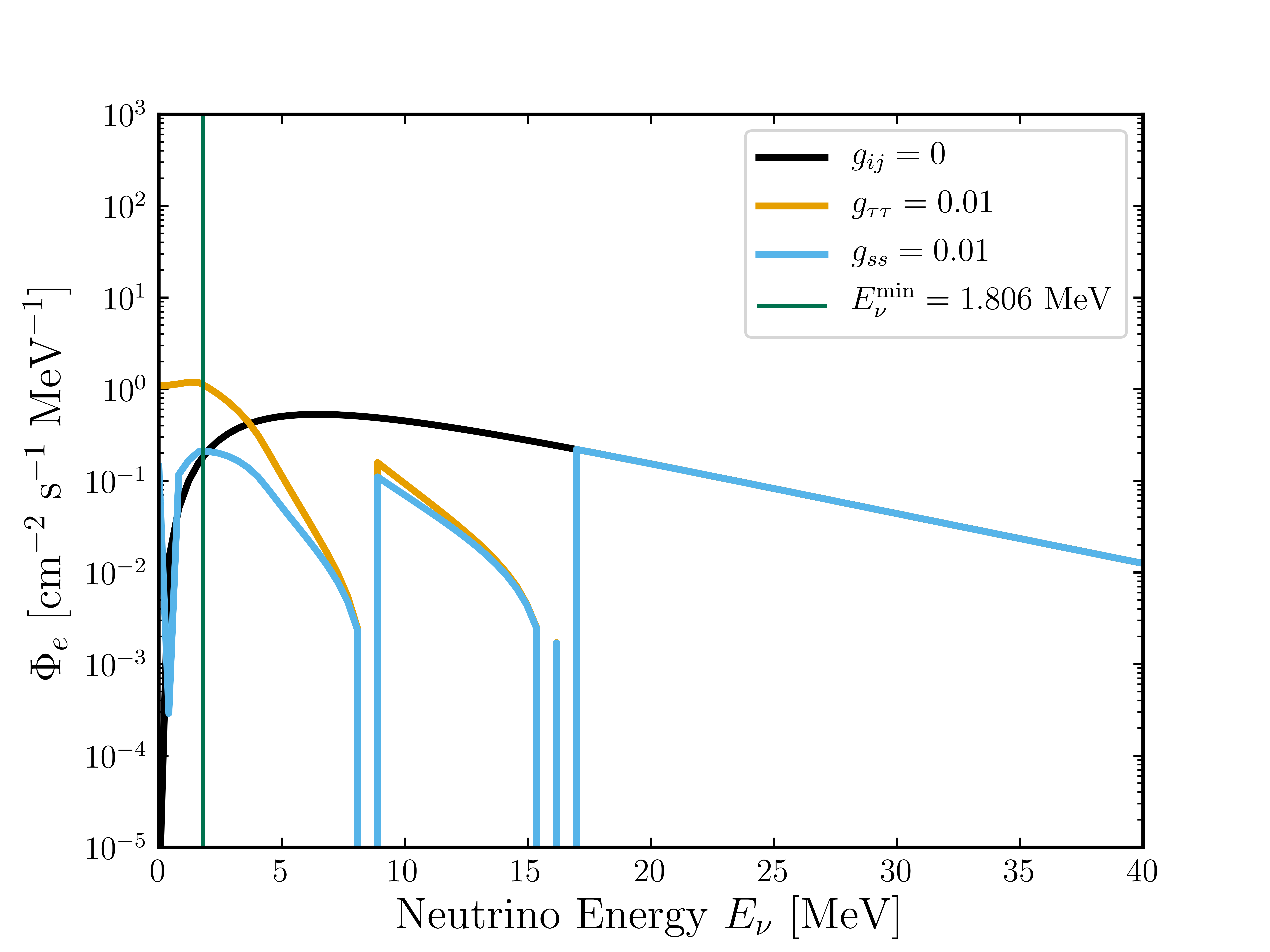}
\caption{The DSNB specific flux of electron anti-neutrinos $\Phi_e$. The forest green line indicates the minimum energy $E_{\nu}^{\rm min} = 1.806\ {\rm MeV}$ needed for neutrinos to undergo inverse beta decay. The black line $g_{ij} = 0$ has no self-interactions. For $\tau$ self-interactions $g_{\tau\tau}$ three resonances are visible, while for $s$ self-interactions $g_{ss}$ four. In both cases there is a nearly degenerate pair of resonances. In addition to the dips, an enhancement is present for energies $E_\nu \lesssim 4\ {\rm MeV}$ for $g_{\tau\tau} = 0.01$, as there is no dip from a fourth neutrino. The mass spectrum is $\vec{m}_{\rm HNH}$ ($\vec{m}_{\rm SNH}$) for the 3 (4) neutrino model. The mediator mass is $m_\phi c^2 = 1\ {\rm keV}$ and the supernova temperature is $k_B T_{\rm SN} = 8\ {\rm MeV}$.
}\label{fig:DSNB_spec}
\end{figure}

The production rate of neutrinos per comoving area per unit time per unit energy from core-collapse supernovae (CCSN) is $S_{i}(t, E) = c R_{\rm CCSN}(z)dN_i(E)/dE$~\cite{astro-ph/0410061}, with $R_{\rm CCSN}$ the CCSN rate per comoving volume, and $dN_i/dE$ the number spectrum of neutrinos of type $i$ emitted by one supernova explosion. For $R_{\rm CCSN}$ we use the parameterization of Ref.~\cite{0812.3157} with the lower bound of the Salpeter initial mass function. Moreover, we assume equipartition of energy among neutrino species and thus approximate the spectrum of one neutrino species by a Fermi-Dirac distribution with zero chemical potential~\cite{astro-ph/0509456}
\begin{align}
\frac{dN_i}{dE} &= \frac{120}{7\pi^4}\frac{E_{\rm tot}}{6}\frac{E^2}{(k_B T_{\rm SN})^4}\frac{1}{1 + e^{E/(k_B T_{\rm SN})}},
\end{align} 
with $E_{\rm tot} = 3\times 10^{46}\ {\rm J}$ the total energy in neutrinos emitted by the supernova and $4\ {\rm MeV} \leq k_B T_{\rm SN} \leq 8\ {\rm MeV}$ the supernova temperature~\cite{1004.3311}. We plot two possible flux spectra $\Phi_e$ of electron anti-neutrinos from the DSNB in Fig.~\ref{fig:DSNB_spec} for $T$. While a supernova temperature $k_B T_{\rm SN} = 8\ {\rm MeV}$ is disfavored, it does not heavily alter our conclusions. For the heaviest normal neutrino mass hierarchy, three resonances are potentially observable when flavor self-interactions are considered. While not observed yet, the addition of gadolinium sulfate to large water Cerenkov detectors would allow for the discrimination, and thus detection, of DSNB events from spallation and atmospheric neutrino events \cite{hep-ph/0309300, 1908.07551}. 

Electron anti-neutrinos in the DSNB are in the correct energy regime to be detected through inverse beta decay scattering at Super-Kamiokande~\cite{1111.5031}. Specifically, through the process $\bar{\nu}_e p \rightarrow e^+ n$, DSNB anti-neutrinos collide with water molecules in Super-Kamiokande, producing a positron that emits Cherenkov radiation that is detectable. As a result, the colliding anti-neutrino must have minimum energy $E_\nu^{\rm min} = m_e c^2 + \Delta = 1.806\ {\rm MeV}$, with $\Delta = (m_n - m_p)c^2$. In the following, we use Eq.~(25) of Ref.~\cite{astro-ph/0302055} for the inverse beta decay cross section $\sigma_{\rm IBD}$. The number $N_{\rm events}$ of events detected in a positron energy bin $[E_{e^+}, E_{e^+} + \delta E]$ is then 
\begin{align}
N_{\rm events} = T N_p \int_{E_{e^+}}^{E_{e^+} + \delta E}dE \Phi_e(E + \Delta)\sigma_{\rm IBD}(E),
\end{align} 
with $T$ the time of observation and $N_p$ the number of scattering targets. Note that in this expression $E + \Delta$ is the neutrino energy, while $E$ is the positron energy.

\begin{figure}[]
\includegraphics[width = \linewidth]{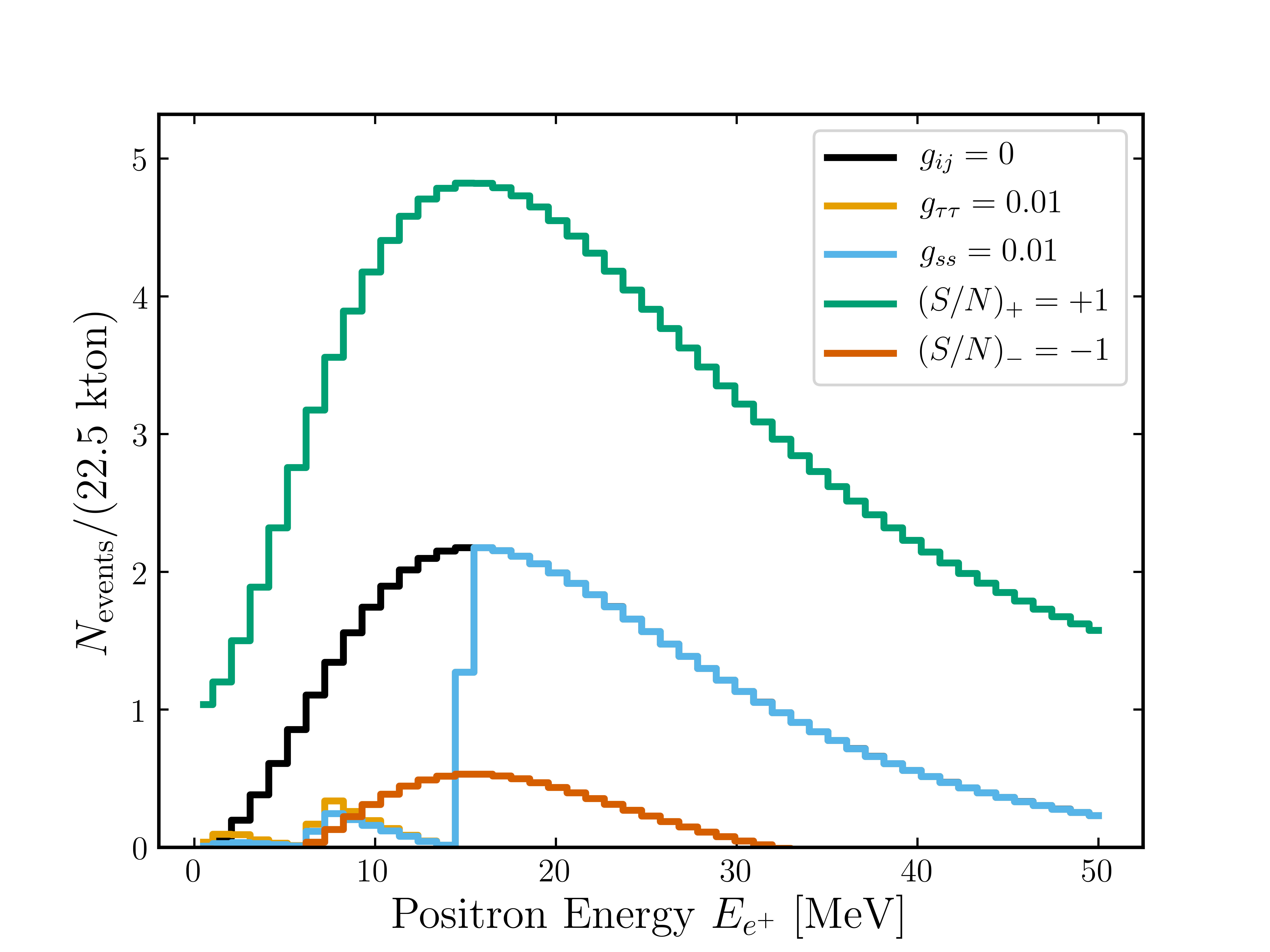}
\caption{DSNB event counts $N_{\rm events}$ vs positron energy $E_{e^+}$ at Super-K with gadolinium after $T = 10\ {\rm years}$ with $\delta E = 1$ MeV energy bins. The upper and lower uncertainties on the $g_{ij}= 0$ event count are shown for $(S/N)_{\pm} = \pm 1$. In both alternative models, self-interactions are ruled out as the resonant energy count is below the $1\sigma$ uncertainty. However, they cannot be distinguished from one another due to their similar profiles. The model parameters are the same as in Fig.~\ref{fig:DSNB_spec}.}\label{fig:SK_hist}
\end{figure}

\begin{figure}[]
\subfloat[]{\includegraphics[width = \linewidth]{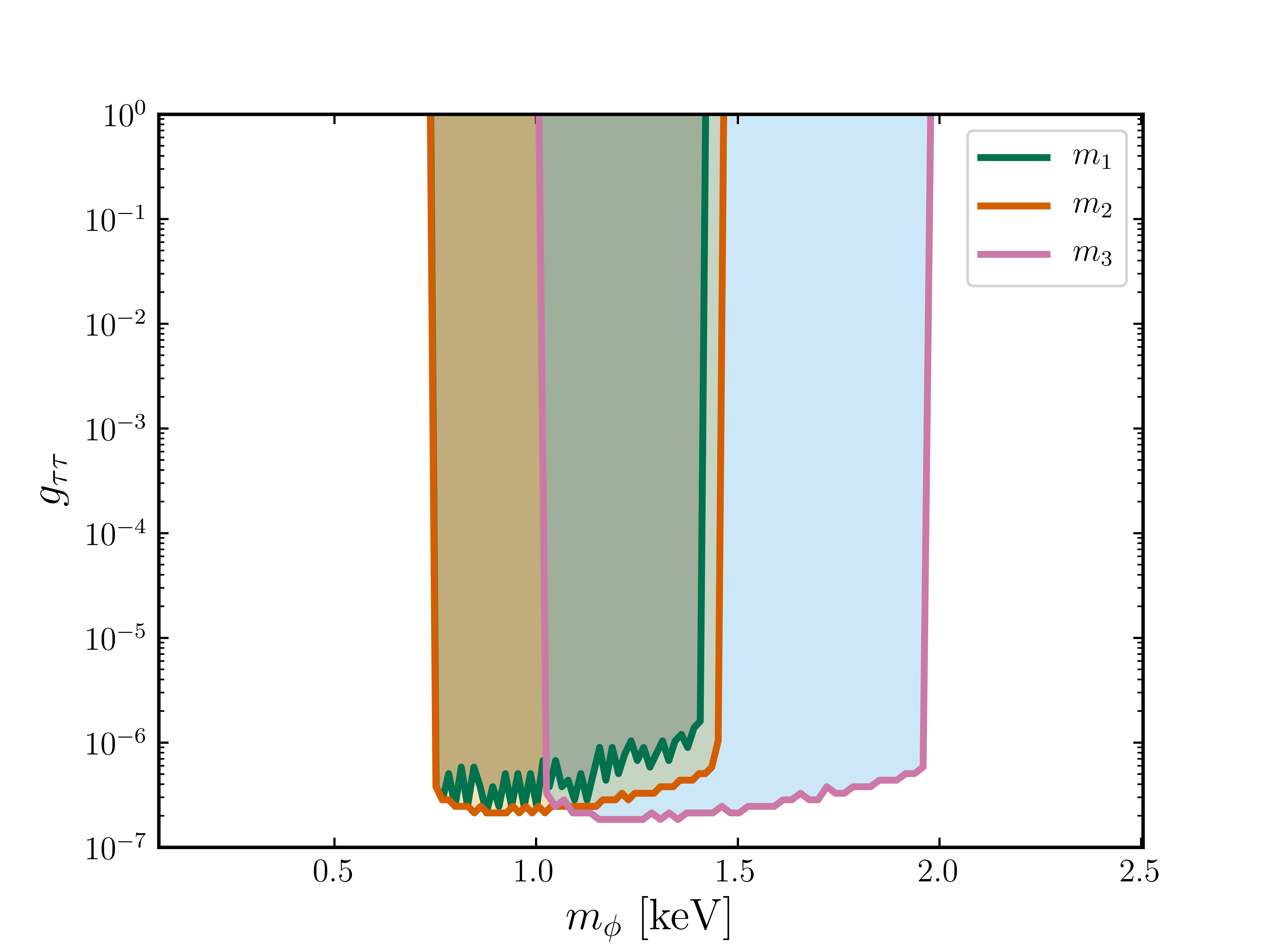}
}
\qquad
\subfloat[]{\includegraphics[width = \linewidth]{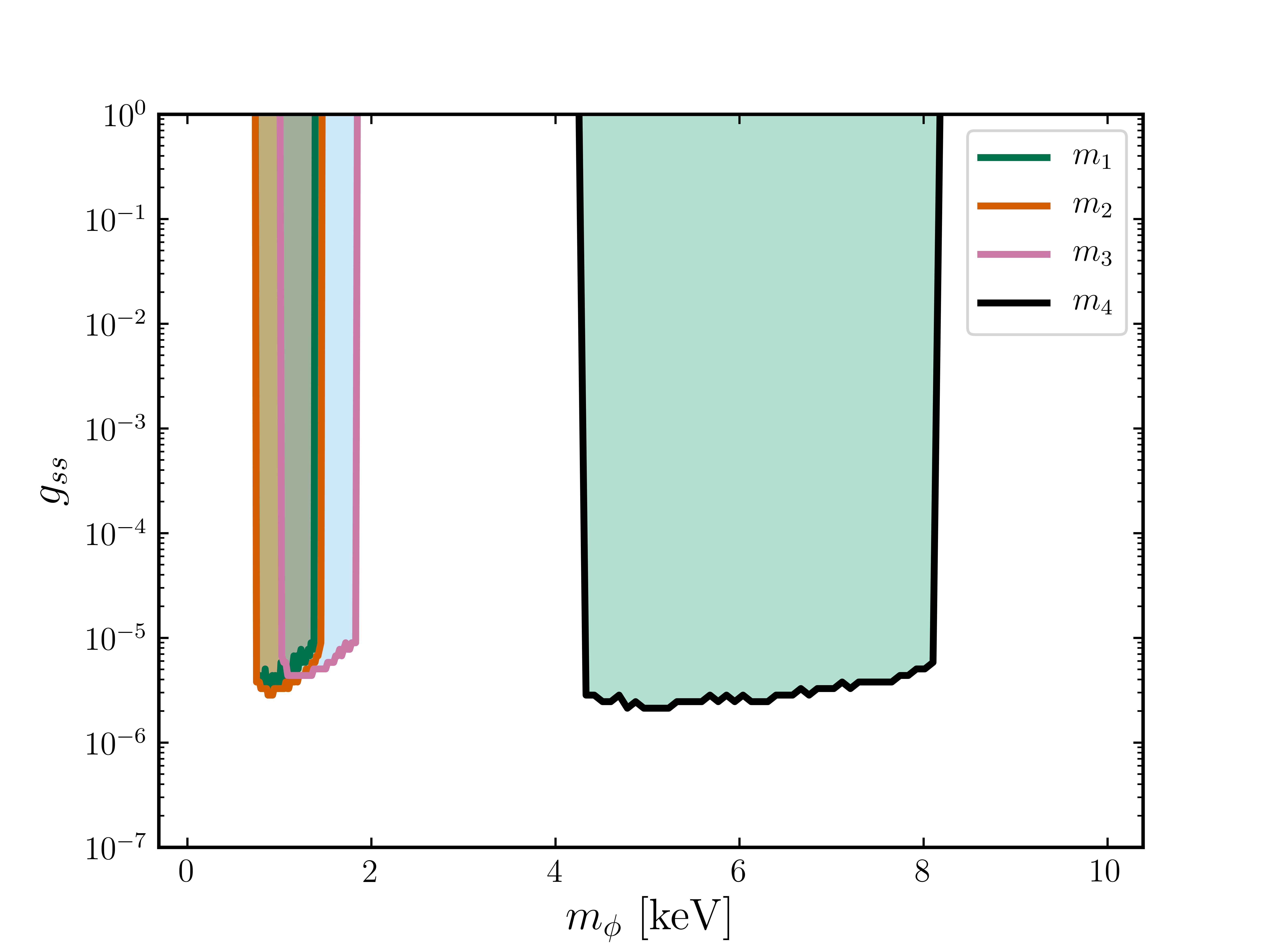}
}
\caption{Forecasted $1\sigma$ constraints on flavor self-interactions from a cosmic neutrino mass spectrum (a) $\vec{m}_{\rm HNH}c^2 = \lbrack 0.030, 0.031, 0.059\rbrack$ eV, (b) $\vec{m}_{\rm SNH}c^2 = \lbrack 0.031, 0.031, 0.059, 1.0\rbrack$ eV interacting with the DSNB observed at Super-K with gadolinium for $T = 10\ {\rm years}$. Each neutrino mass $m_j$ corresponds to a different constraint region, denoted by the filled in regions. The jagged edges are due to numerical error.}\label{fig:SK_constraint}
\end{figure}

We show the event counts and uncertainties corresponding to Fig.~\ref{fig:DSNB_spec} in Fig.~\ref{fig:SK_hist}.  
Comparing our null hypothesis to our model at the resonant energies, we obtain the forecasted $1\sigma$ constraints in Fig.~\ref{fig:SK_constraint}.

\subsection{High-Energy Astrophysical Neutrinos}\label{subsec:IceCube}
\begin{figure}[]
\includegraphics[width = \linewidth]{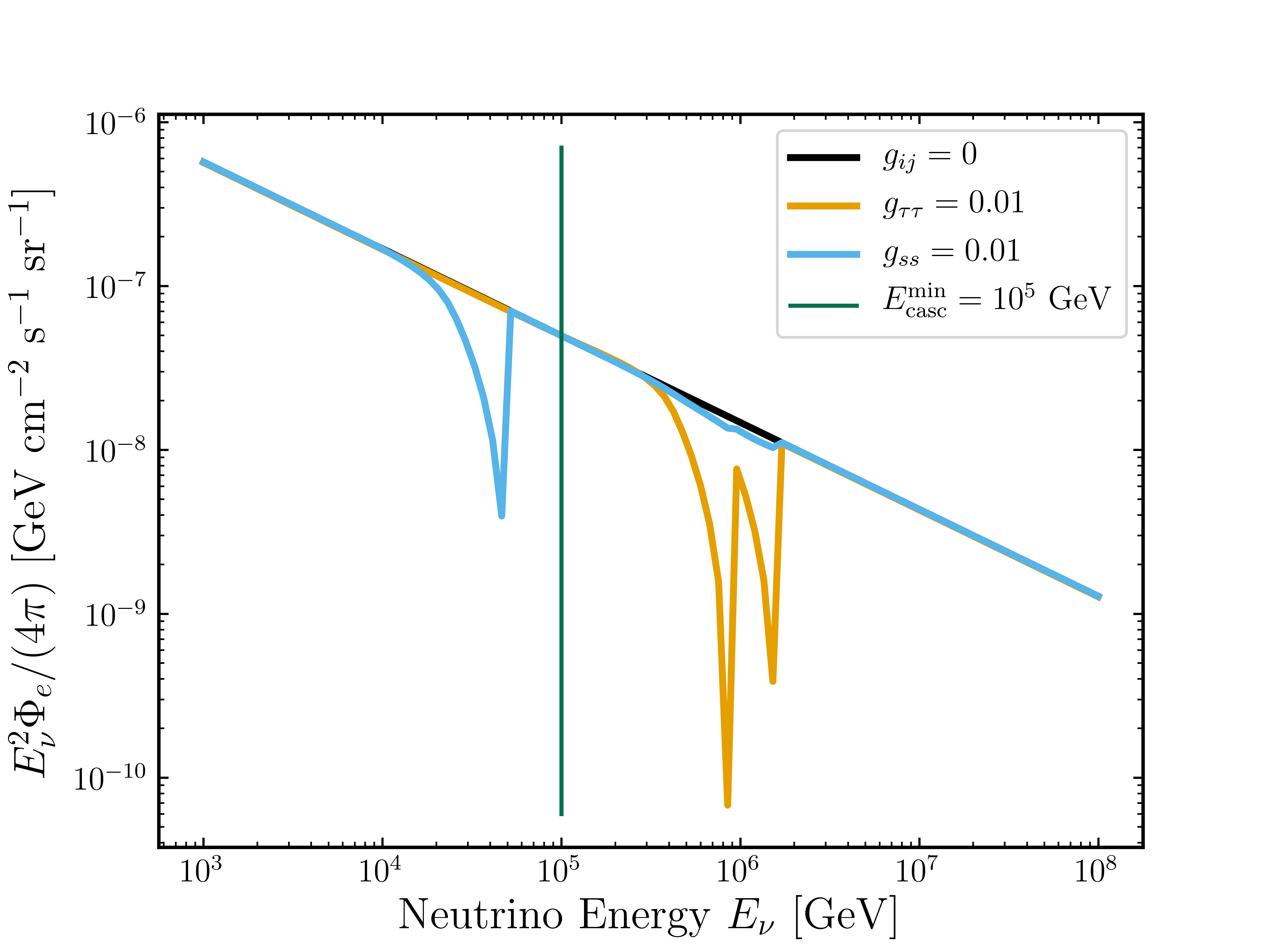}
\caption{The specific flux per steradian $\Phi_e/(4\pi)$ of high-energy astrophysical electron anti-neutrinos. The forest green line indicates the minimum energy $E^{\rm min}_{\rm casc } = 10^5\ {\rm GeV}$ needed to be above the atmospheric neutrino background. The black line $g_{ij} = 0$ has no self-interactions. For the $\tau$ self-interactions $g_{\tau\tau}$ two resonances are visible, while for $s$ self-interactions $g_{ss}$ one. In the $g_{\tau\tau}$ case there is a degenerate pair of resonances that cannot be resolved. For $g_{ss} = 0.01$ three resonances are below the threshold for strong absorption. No enhancement is present for low energies as the spectrum monotonically decays. The mass spectrum is $\vec{m}_{\rm HNH}$ ($\vec{m}_{\rm SNH}$) for the 3 (4) neutrino model. The mediator mass is $m_\phi c^2 = 10\ {\rm MeV}$. We take the power law index to be $\gamma = 2.53$ and $E_0 = 100$ TeV. In addition, we normalize the final flux at energy $E_0$ so that $E_0^2 \Phi_e(E_0)/(4\pi) = C_0 \Phi_0$, with $C_0 = 3\times 10^{-18}\ {\rm GeV}^{-1}{\rm cm}^{-2}{\rm s}^{-1}{\rm sr}^{-1}$ and $\Phi_0 = 1.66$.
}\label{fig:HEAN_spec}
\end{figure}

The production rate of high-energy astrophysical neutrinos per comoving area per unit time per unit energy is $\mathcal{L}(z, E) = \mathcal{W}(z)\mathcal{L}_0(E)$, with $\mathcal{L}_0$ the differential number luminosity for each source and $\mathcal{W}$ the redshift evolution of the source density. We take the redshift evolution to follow the star-formation rate, $\mathcal{W}(z) = R_{\rm CCSN}(z)$. Moreover, following IceCube's 6-year data analysis~\cite{2001.09520}, we take the differential number luminosity to be a power law $\mathcal{L}_0 \propto (E/E_0)^{-\gamma}$. We plot two possible flux spectra $\Phi_e$ of electron anti-neutrinos in Fig.~\ref{fig:HEAN_spec}. The number of events observed by IceCube is~\cite{1306.2309} 
\begin{align}
N_{\rm events} = T\int_{E_{\rm casc}}^{E_{\rm casc} + \delta E}dE \Phi_e(E) A_{\rm eff}(E),
\end{align}
with $T$ the time of observation and $A_{\rm eff}(E)$ the IceCube effective area, which we take from Ref.~\cite{1311.5238}. Note that we are approximating the neutrino energy to be the cascade energy.
 
\begin{figure}[h!!]
\includegraphics[width = \linewidth]{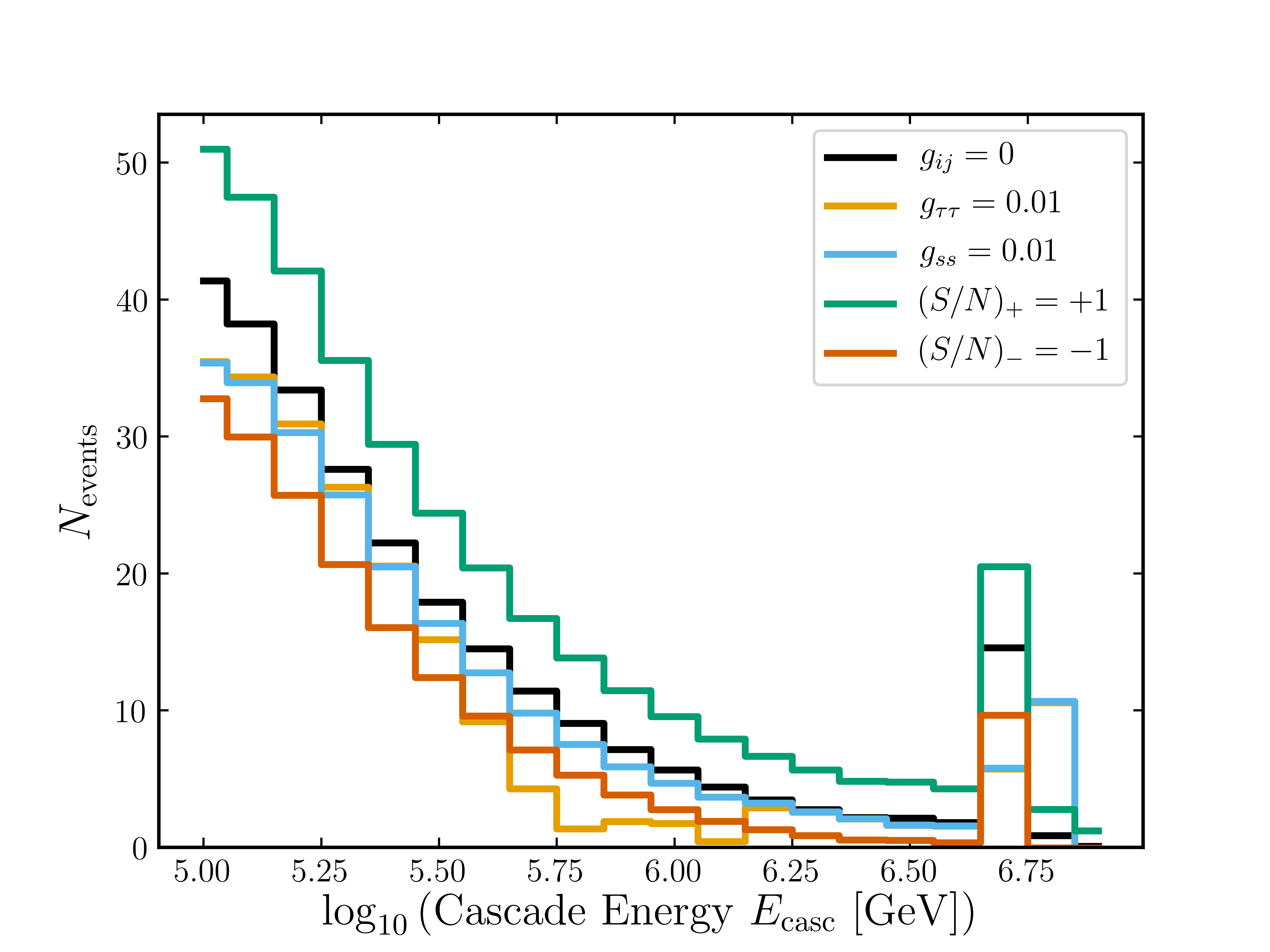}
\caption{High-energy astrophysical neutrino event counts $N_{\rm events}$ vs cascade energy $E_{\rm casc}$ at IceCube after $T = 988\ {\rm days}$ with $\delta \log_{10} \lbrack E/(1\ {\rm GeV})\rbrack = 0.1$ log-energy bins. The upper and lower uncertainties on the $g_{ij}= 0$ event count are shown for $(S/N)_{\pm} = \pm 1$. The $g_{\tau\tau} = 0.01$ self-interaction model is ruled out as the resonant energy count is below the $1\sigma$ uncertainty. The model parameters are the same as in Fig.~\ref{fig:HEAN_spec}.}\label{fig:IC_hist}
\end{figure}

\begin{figure}[]
\subfloat[]{\includegraphics[width = \linewidth]{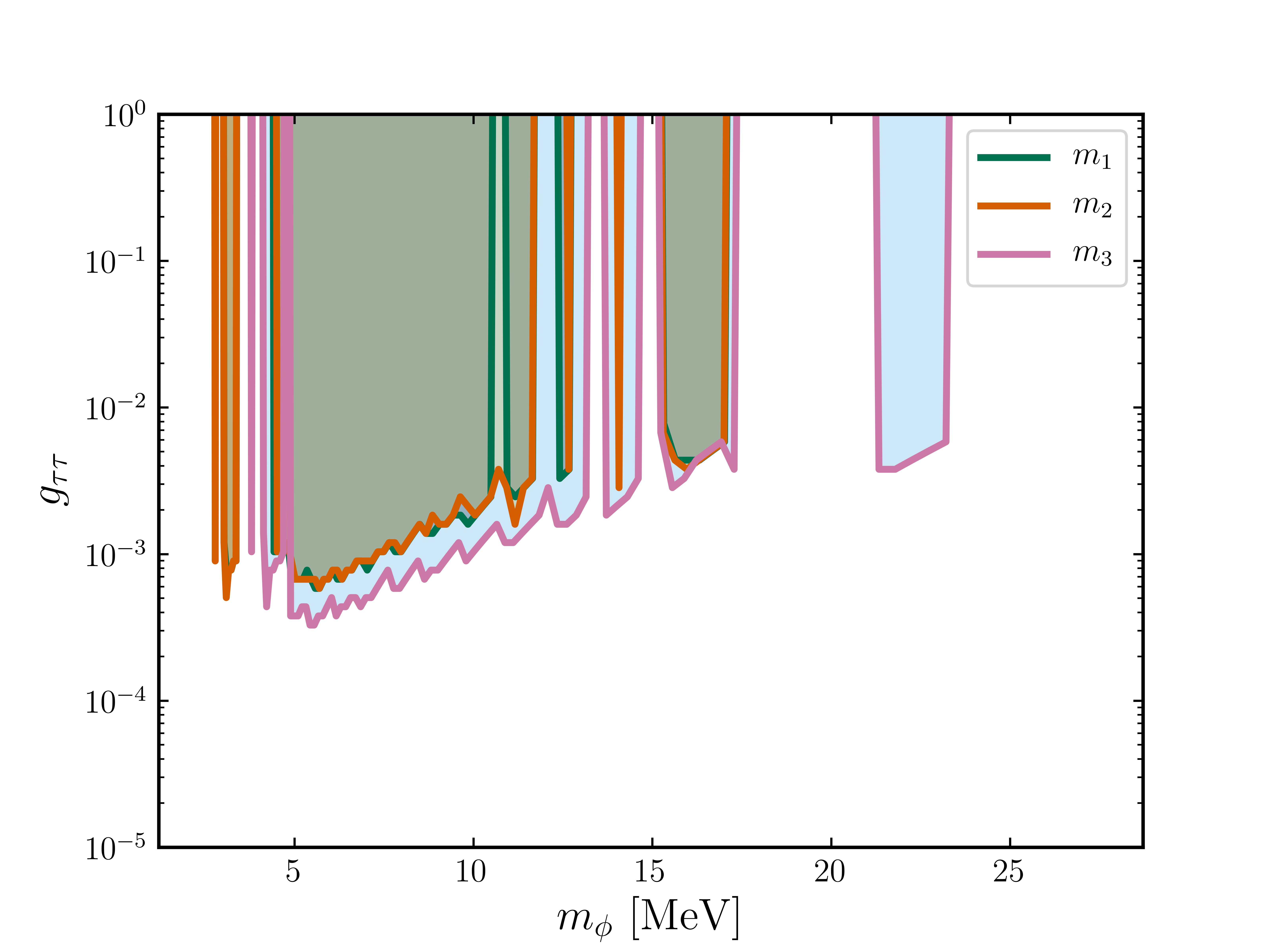}
}
\qquad
\subfloat[]{\includegraphics[width = \linewidth]{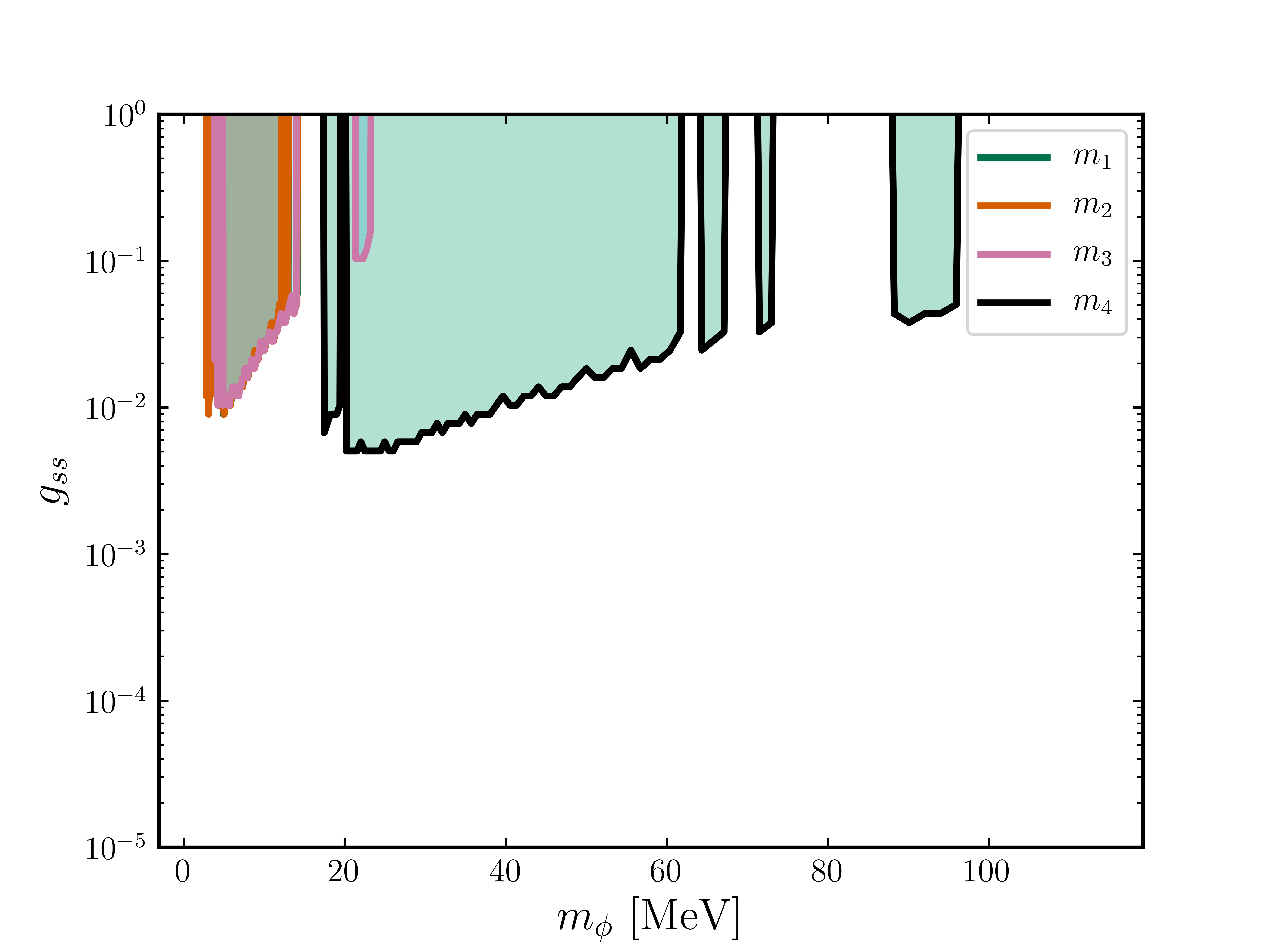}
}
\caption{Forecasted $1\sigma$ constraints on flavor self-interactions from a cosmic neutrino mass spectrum (a) $\vec{m}c^2 = \lbrack 0.030, 0.031, 0.059\rbrack$ eV, (b) $\vec{m}c^2 = \lbrack 0.031, 0.031, 0.059, 1.0\rbrack$ eV (b) interacting with HEAN observed at IceCube for $T = 988\ {\rm days}$. Each neutrino mass $m_j$ corresponds to a different constraint region, denoted by the filled in regions. The jagged edges are due to numerical error.}\label{fig:IC_constraint}
\end{figure}

We show the event counts and uncertainties corresponding to Fig.~\ref{fig:HEAN_spec} in Fig.~\ref{fig:IC_hist}.
Comparing our null hypothesis to our model at the resonant energies, we obtain the forecasted $1\sigma$ constraints in Fig.~\ref{fig:IC_constraint}.
\section{Discussion}\label{sec:disc}
Several points are worth examining in further depth. First, Eq.~\eqref{eq:gen_flux} only holds when each cosmic neutrino species is cold. However, we know there exists at least one cold neutrino species. Therefore, in the case where one or more cosmic neutrino species are not cold, this equation is modified so that any sum over neutrino scattering cross sections is only over all cold species.  Moreover, interactions with the lower-mass neutrinos should be suppressed relative to that of the heavier cold species due to thermal broadening.  

Second, the spectra shown all have three resonances, and this not need be the case. The heaviest allowed normal neutrino mass hierarchy is special in this case, since the nearly-degenerate pair have masses much larger than any neutrino mass splitting. In the inverted scenario, this cannot be the case and so at most two resonances could be seen in any spectra that does not have a large energy range. If a single resonance is seen it is unclear how to disentangle the two scenarios, but such distinction is outside the scope of this work. 

Third, while we made our constraints by looking for absorption features, one in principle could also look for enhancements in spectra. In experiments, it is simple as one needs to look for when the signal surpasses some threshold for statistical significance, $N_{\rm events} > N_+$. However, it is less obvious theoretically what bins or how many bins one should look at in order to create a forecasted constraint from enhancements in a time-efficient manner. It depends on the number of resonances, the shape of the null hypothesis spectrum, and the detection method. For example, in the DSNB the enhancements are much more pronounced at low energy compared to HEAN sources, since at low energies the DSNB spectrum falls off while the HEAN source spectrum grows. 
 
Fourth, we wrote down our formulas assuming a single scalar $\phi$, however it is straightforward to generalize to multiple scalars $\phi_k$ with self-coupling matrices $g^k_{ij}$. The only possible subtlety is if degeneracies in the resonances occur, in which case the resonant condition needs to be altered accordingly by a sum over degenerate resonances.

Fifth, while this paper is focused on neutrino self-interactions, it is also straightforward to incorporate arbitrary resonant scattering between any species. The most obvious other cold species to generalize to would be cold dark matter.

Finally, we took the noise to be only Poissonian and assumed fiducial astrophysical parameters. In a realistic experiment, other backgrounds must be taken into account as well as degeneracies with their parameters. However, such a proper treatment, similar to Refs.~\cite{1205.6292, 1804.03157},  is outside the scope of this work. 

\section{Conclusion}\label{sec:conc}
In this paper we have have considered the consequence of beyond the Standard Model neutrino self-interactions on various astrophysical neutrino spectra. We began by presenting the necessary formalism for neutrino mixing and transport. We did this not only to establish notation, but also in order to demonstrate that neutrino reinjection is a problem that is generally not closed. 

In order to overcome this hurdle, we then took the limit where the scattering cross section goes to a delta function, and found that the former partial integro-differential equations turn into a standard partial differential equation with simple boundary conditions. As a result, we then presented the solution for astrophysical neutrino spectra for a single neutrino species, following with one for an arbitrary number of neutrino species. These solutions were specified in either the mass basis or the flavor basis. 

From this, we then demonstrated the utility of the analytic solution by considering our astrophysical sources to be either the diffuse supernovae background or high-energy astrophysical neutrinos. From there, we established forecasts and constraints on a normal 3-neutrino hierarchy with $\tau$ self-interactions, as well as a 4-neutrino hierarchy with sterile self-interactions. None of these calculations took a significant amount of time, and were routine in their implementation. 

It will be interesting to implement this calculation in future work to explore the effects of neutrino self-interactions on DSNB and HEAN spectra for a wider range of models that involve new neutrino interactions. 

\subsection*{Acknowledgments}
We thank Bei Zhou and John Beacom for useful discussions. CCS Acknowledges the support of the Bill and Melinda Gates Foundation. This work was supported at Johns Hopkins by NASA Grant No. NNX17AK38G, NSF Grant No. 1818899, NSF Grant No. DGE-1746891, and the Simons Foundation.

\appendix
\section{Flavor Basis Interactions}\label{ap:flavor_int}
We consider the most general flavor interaction for a single scalar mediator $\phi$ of mass $m_\phi$
\begin{align}
\nonumber \mathcal{L}_{\rm int}^{\rm flavor} &= \phi\sum_{\alpha \beta}g_{\alpha \beta}\nu_{\alpha}\nu_\beta\\
&= \phi\sum_{\alpha\beta i j}g_{\alpha\beta}U_{\alpha i}U_{\beta j}\nu_i \nu_j.
\end{align}
The identification of $g_{ij} = \sum_{\alpha\beta} U_{i\alpha}U_{j\beta}g_{\alpha\beta}$ allows us to use Eq.~\eqref{eq:gen_flux}. We reparameterize the scattering rate 
\begin{align}\Gamma_R^{jkil} = \sum_{\alpha\beta\gamma\delta}|U_{\gamma i}|^2 |U_{\alpha j}|^2 |U_{\delta l}|^2 \Gamma_R^{k, \alpha\beta\gamma\delta},
\end{align}
with $\Gamma_R^{k, \alpha\beta\gamma\delta} = |U_{\beta k}|^2 n_k(t) \sigma^{\alpha\beta\gamma\delta}_R c$ and $\sigma^{\alpha\beta\gamma\delta}_R = |g_{\alpha\beta}|^2|g_{\gamma\delta}|^2/[4(m_\phi c^2)\Gamma_\phi]$. We choose such a reparameterization in order to separate the neutrino conversion probabilities from the scattering cross sections. In doing so, and invoking unitarity, we obtain the result
\begin{align}
\nonumber \Phi_i(t, E) &= \int_{-\infty}^t dt'[a(t)/a(t')]e^{-\tau_i(t', t, E)}\widetilde{S}_i\{t', [a(t)/a(t')E]\},\\
\nonumber \widetilde{S}_i(t, E) &= S_{i}(t, E) + \sum_{\gamma\delta}|U_{\gamma i}|^2\left(1  + |U_{\delta i}|^2\right)\\
\nonumber\hphantom{\widetilde{S}_i(t, E)}&\hphantom{=S_i(t, E) + } \times \sum_{\alpha\beta k} \Gamma_R^{k,\alpha\beta\gamma\delta}(t)\Phi_\alpha(t, E_k)\Theta(E_k - E),\\
\nonumber \Phi_i(t, E_j) &= \frac{H(t)}{\Gamma^{ij}_R(t)}\int_{-\infty}^t dt'\frac{a(t)}{a(t')}\left[e^{-\tau_i(t', t, E_j^+)} - e^{-\tau_i(t', t, E_j^-)}\right]\\
\hphantom{\Phi_i(t, E_j)}&\hphantom{= \frac{H(t)}{\Gamma^{ij}_R(t)}\int_{-\infty}^t dt'\ \frac{a(t)}{a(t')}}\times \widetilde{S}_i\{t', [a(t)/a(t')]E_j\},
\end{align}
with $\Gamma^{ij}_R(t) = \sum_{\alpha}|U_{\alpha i }|^2\left(\sum_{\beta\gamma\delta}\Gamma_R^{j, \alpha\beta\gamma\delta}\right)$. While we have presented here these equations in the flavor basis for analytic insight, we note that in general it is easier numerically to use Eq.~\eqref{eq:gen_flux} with the appropriate substitution, as it contains less summations. 

However, for a single-flavor $\alpha$ interaction, where the flavor self-coupling matrix is $g_{\mu\nu} = g \delta_{\mu\alpha}\delta_{\nu\alpha}$, there is a decidedly simpler form in the flavor basis 
\begin{align}
\nonumber \Phi_i(t, E) &= \int_{-\infty}^t dt'[a(t)/a(t')]e^{-\tau_i(t', t, E)}\widetilde{S}_i\{t', [a(t)/a(t')E]\},\\
\nonumber \widetilde{S}_i(t, E) &= S_{i}(t, E) + |U_{\alpha i}|^2\left(1  + |U_{\alpha i}|^2\right)\\
\nonumber \hphantom{\widetilde{S}_i(t, E)}&\hphantom{= S_i(t, E)  + }\times\sum_k\Gamma_R^{k, \alpha}(t)\Phi_{\alpha}(t, E_k)\Theta(E_k - E),\\
\nonumber \Phi_\alpha(t, E_j) &= \frac{H(t)}{\Gamma^{j, \alpha}_R(t)}\int_{-\infty}^t dt'\frac{a(t)}{a(t')}\left[e^{-\tau_i(t', t, E_j^+)} - e^{-\tau_i(t', t, E_j^-)}\right]\\
\hphantom{\Phi_\alpha(t, E_j)}&\hphantom{=\frac{H(t)}{\Gamma^{\alpha j}_R(t)}\int_{-\infty}^t dt'\ \frac{a(t)}{a(t')}}\times \widetilde{S}_i\{t', [a(t)/a(t')]E_j\},
\end{align}
with $\Gamma^{j, \alpha}_R(t) = |U_{\alpha k}|^2 n_k(t) \sigma_R c$ and $\sigma_R = (\hbar c)^2 \pi g^2/(m_\phi c^2)^2$.

\end{document}